\documentclass[nofootinbib, amsmath,amssymb, prd]{revtex4}
\usepackage{amsmath,amssymb,graphicx}
\usepackage{epsf}
\usepackage{epstopdf}
\usepackage {amssymb}
\usepackage{xcolor}
\newcommand{\nc}{\newcommand}
\nc{\ba}{\begin{eqnarray}}
\nc{\ea}{\end{eqnarray}}
\newcommand\Sm{{{\cal S}_m}}
\newcommand\be{\begin{equation}}
\newcommand\ee{\end{equation}}

\newcommand\mPl{{M_{\rm Pl}}}

\newcommand{\bfk}{{\bf{k}}}

\newcommand{\calP}{{\cal{P}}}

\newcommand{\calS}{{{\cal{S}}_m}}
\newcommand{\calA}{{\cal{A}}}
\newcommand{\calB}{{\cal{B}}}
\newcommand{\calC}{{\cal{C}}}
\nc{\x}{{\bf{x}}}
\nc{\fnl}[1]{{f^{#1}_{NL}}}
\begin{document}
\title{ CMB hemispherical asymmetry from non-linear isocurvature perturbations}

\author{Hooshyar Assadullahi$^{1, 2}$}
\email{hooshyar.assadullahi-AT-port.ac.uk}
\author{Hassan Firouzjahi$^{3}$}
\email{firouz-AT-mail.ipm.ir}
\author{Mohammad Hossein Namjoo$^{4}$}
\email{mohammadhossein.namjoo-AT-utdallas.edu}
\author{David Wands$^{1}$}
\email{david.wands-AT-port.ac.uk}
\affiliation{$^1$ Institute of Cosmology and Gravitation, University of Portsmouth, Dennis Sciama Building, Burnaby Road, Portsmouth PO1 3FX, United Kingdom}
\affiliation{$^2$ School of Earth and Environmental Sciences, University of Portsmouth, Burnaby Building, Burnaby Road, Portsmouth PO1 3QL, United Kingdom}
\affiliation{$^3$School of Astronomy, Institute for Research in
Fundamental Sciences (IPM),
P.~O.~Box 19395-5531,
Tehran, Iran}
\affiliation{$^4$ Department of Physics, The University of Texas at Dallas, Richardson, TX 75083, USA}

\begin{abstract}
\vspace{0.3cm}
We investigate whether non-adiabatic perturbations from inflation could produce an asymmetric distribution of temperature anisotropies on large angular scales in the cosmic microwave background (CMB). We use a generalised non-linear $\delta N$  formalism to calculate the non-Gaussianity of the primordial density and isocurvature perturbations due to the presence of non-adiabatic, but approximately scale-invariant field fluctuations during multi-field inflation. This local-type non-Gaussianity leads to a correlation between very long wavelength inhomogeneities, larger than our observable horizon, and smaller scale fluctuations in the radiation and matter density. Matter isocurvature perturbations contribute primarily to low CMB multipoles and hence can lead to a hemispherical asymmetry on large angular scales, with  negligible asymmetry on smaller scales. 
In curvaton models, where the matter isocurvature perturbation is partly correlated with the primordial density perturbation, we are unable to obtain a significant asymmetry on large angular scales while respecting current observational constraints on the observed quadrupole. However in the axion model, where the matter isocurvature and primordial density perturbations are uncorrelated, we find it may be possible to obtain a significant asymmetry due to isocurvature modes on large angular scales. 
Such an isocurvature origin for the hemispherical asymmetry would naturally give rise to a distinctive asymmetry in the CMB polarisation on large scales.
\vspace{0.3cm}
\end{abstract}

\date\today


\maketitle

\section{Introduction}

The standard $\Lambda$CDM cosmology provides a remarkably good fit to current observations of the cosmic microwave background temperature anisotropies and large-scale structure in high-redshift galaxy surveys. Nonetheless there are hints of anomalies on the largest observable scales which appear to be hard to reconcile with a simple, almost scale-invariant, Gaussian distribution of primordial density perturbations as expected due to quantum fluctuation in simple inflationary models for the origin of structure.

One of the most intriguing results from cosmic microwave background (CMB) satellite experiments is the suggestion of a weak hemispherical asymmetry in the temperature fluctuations \cite{Ade:2013nlj, Eriksen:2003db,Hansen:2004vq,Eriksen:2007pc,Hansen:2008ym,Hoftuft:2009rq, Gordon:2005ai, 
Akrami:2014eta,Rath2013, Adhikari:2014mua, Quartin:2014yaa, Notari:2013iva}.  The dipole statistical anisotrpoy can be modeled by the following relation \cite{Ade:2013nlj,Gordon:2005ai}
\ba
\label{dT-anisotropy}
\frac{\delta T}{T}(\hat{n})=\left(\frac{\delta T}{T}\right)^0(\hat{n})\left[1+A^0\hat{n}.\hat{p}\right]
\ea
where $(\frac{\delta T}{T})^0(\hat{n})$ is the isotropic part, $\hat{n}$ is the direction in the sky, $A^0$ is the amplitude of the hemispherical asymmetry and $\hat{p}$ is the preferred direction. The Planck collaboration finds the amplitude $A=0.073\pm0.010$ for the scales corresponding to $\ell=2-64$ and the preferred direction $(217.5\pm15.4 , -20.2\pm 15.1)$ in galactic coordinates \cite{Ade:2013nlj}.

Although at first sight this appears in conflict with homogeneous and isotropic distribution of perturbations, it could be due to a very large scale perturbation (on scales larger than our present Hubble horizon) leading to a gradient across our observable universe \cite{kam1, kam2, pesky, lyth, Wang2013, Lou2013, Namjoo:2013, Liddle2013,
Mazumdar2013, Abolhasani:2013, jm1, jm2, Kanno2013, Firouzjahi:2014mwa,McDonald:2014lea,McDonald:2014kia, Cai2013} (See e.g. \cite{Amico2013,Kohri2013,Jazayeri:2014nya,Liu2013}
 for other proposals.). If temperature fluctuations on our CMB sky are correlated with this very large scale mode then it could produce a hemispherical asymmetry in our observable universe. That is, a very large scale fluctuation could spontaneously break isotropy in our observable patch while remaining part of an isotropic distribution on far larger scales. 

Another potential challenge to the simplest models of an isotropic and almost scale invariant primordial power spectrum comes from the surprisingly low power observed in the lowest multipoles in the CMB sky. The observed temperature fluctuations for multipole numbers $\ell<40$ appear to lie systematically below the $\Lambda$CDM predictions for an adiabatic density perturbations. Several solutions have been proposed including broken scale-invariance, suppressing large-scale perturbations, or running of the spectral index, or isocurvature matter perturbations (anti-)correlated with the radiation density perturbations, which can lead to a cancellation in the large-scale temperature fluctuations \cite{Langlois:1999dw,Bucher:1999re,Amendola:2001ni}.

Either or both of these anomalies may point to the role of non-adiabatic perturbations in the very early universe. 
A local-type non-Gaussianity in real space \cite{Wands:2010af} leads to correlations between long and short wavelength modes in Fourier space (corresponding to a non-zero bispectrum in squeezed configurations) of the sort which might produce the hemispherical asymmetry on the CMB sky from very long wavelength perturbations. However local-type non-Gaussianity is strongly suppressed in simple models of inflation that rely on a single inflaton field to drive inflation and produce primordial density perturbations from fluctuations in that field \cite{Maldacena:2002vr,Acquaviva:2002ud,Creminelli:2004yq}. Local-type non-Gaussianity can only be produced by non-linear evolution from non-adiabatic field perturbations during inflation, requiring multiple light fields during slow-roll inflation. These non-adiabatic field perturbations can also give rise to isocurvature matter perturbations depending upon the subsequent evolution after inflation, which may or may not be correlated with the primordial curvature perturbation \cite{Bassett:2005xm}.

There is no evidence of any dipole asymmetry in CMB fluctuations on smaller angular scales or in tracers of large-scale structure, which leads to tight constraints on a simple dipole asymmetry in the primordial matter perturbations on smaller comoving scales. Quasar observations put a lower bound $A<0.012$ at 95\% confidence \cite{Hirata:2009ar}. Also a recent analysis with Planck data put the limit of $A<0.0045$ for the moments of $\ell=601-2048$ \cite{Flender:2013jja}. This again suggests the possibility that the asymmetry is associated with isocurvature perturbations since the linear transfer function for isocurvature perturbations is suppressed on small scales with respect to the transfer function for adiabatic perturbations.

In this paper we present general expressions for curvature and isocurvature perturbations due to the evolution of adiabatic and entropy field fluctuations during inflation and show how the non-linear evolution associated with primordial non-Gaussianities can lead to hemispherical asymmetries in the CMB on large angular scales from both curvature and isocurvature perturbations. The physical intuition behind this effect is simple: If a long wavelength super horizon mode is responsible for asymmetry on small scale (CMB) modes, it must be due to non-adiabatic fluctuations and non-linear evolution which correlates modes at different scales. The larger the non-linearity, the stronger the correlations between different scales, and hence the larger is the amplitude of asymmetry.  

The idea of generating dipole asymmetry from isocurvature mode is originally presented in \cite{Erickcek:2009at} in which the authors only considered the curvaton model. We extend the proposal to a model independent formalism so that it can be applied to any model of isocurvature. One of the advantages of our formalism is that the relation between non-linearity and the dipole asymmetry is manifest. We examine the formalism by applying it to the curvaton model and then, for the first time, we also study the axion model as another possibility for generating the dipole asymmetry. Furthermore, we obtain model independent constraints on the amplitude of the long wavelength gradient mode  from low $\ell$ CMB multipoles and update these constraints using recent Planck data.  

It is important to recognise that we only have limited evidence for very large scale anomalies in our universe if these are to be interpreted as random fluctuations of an underlying Gaussian distribution. We have relatively few independent measurements of the distribution on scales approaching the observed volume, and thus the expected cosmic variance is significant, diminishing the statistical significance of deviations from the expected value. This can be partially addressed by seeking additional independent measurements, such as the polarisation as well as intensity of the CMB anisotropy. We discuss possible tests of non-linear and non-adiabatic perturbations leading to hemispherical asymmetry in our conclusions.

\section{The non-adiabatic $\delta$N-formalism}

We will assume that the observed primordial density perturbation constrained by current observations arises from quantum vacuum fluctuations in light scalar fields during inflation in the very early universe. 
Primordial adiabatic density or curvature perturbations (e.g., at the epoch of primordial nucleosynthesis) can be identified with the perturbed expansion, $N$, from an initial spatially flat hypersurface during inflation to a uniform total density hypersurface in the primordial (radiation-dominated) era, due to initial field fluctuations during inflation \cite{Sasaki:1995aw, Lyth:2004gb, Lyth:2005fi}
\be
\zeta = \sum_A \frac{\partial N}{\partial\varphi^A} \delta\varphi^A + \frac12 \sum_{A,B} \frac{\partial^2 N}{\partial\varphi^A\partial\varphi^B} \delta\varphi^A \delta\varphi^B + \ldots \,.
\ee
Primordial matter isocurvature perturbations correspond to a relative perturbation between the non-relativistic matter and radiation, hence we can identify this as the difference in the perturbed expansion, $\Delta N$, between a uniform total matter density hypersurface and a uniform total radiation density hypersurface\footnote{The factor of 3 is conventional, so that at linear order $\Sm$ corresponds to the fractional matter density perturbation $\delta\rho_m/\rho_m$ on uniform radiation density hypersurfaces.} \cite{ Langlois:2008vk}
\be
\Sm = 3\left( \sum_A \frac{\partial \Delta N}{\partial\varphi^A} \delta\varphi^A + \frac12 \sum_{A,B} \frac{\partial^2 \Delta N}{\partial\varphi^A\partial\varphi^B} \delta\varphi^A \delta\varphi^B + \ldots \right) \,.
\ee

It is convenient to write the field fluctuations during inflation, $\delta\varphi^A$, in terms of {\em adiabatic} field fluctuations tangential to the background trajectory in field space, $\delta\sigma$, and {\em entropy} fluctuations orthogonal to the background trajectory \cite{Gordon:2000hv}. For simplicity we consider only a two-field, canonical slow-roll model with one entropy field direction, but it is straightforward to generalise to include more fields. 
This allows us to simplify the expressions for the curvature and isocurvature perturbations since the curvature perturbation, $\zeta$, receives contributions from both adiabatic and entropy perturbations whereas the matter isocurvature perturbation, $\Sm$, will only be sourced by entropy field perturbations. Following the notation in \cite{Langlois:2011zz} we write the curvature and iso-curvature perturbations by
\ba
\label{def_primordial}
\zeta &=& \zeta_i + z_1 \hat S + \dfrac{1}{2} z_2 \, \hat S^2 
 \nonumber \\
\Sm &=& s_1 \hat S+ \dfrac{1}{2} s_2 \, \hat S^2.
\ea
Note that the curvature perturbation due to adiabatic field perturbations during inflation, $\zeta_i$, is well-described by a Gaussian distribution. Second-order local-type non-Gaussianity from adiabatic field perturbations is proportional to the scale-dependence of the power spectrum, $f_{NL}\sim n-1$ \cite{Maldacena:2002vr,Creminelli:2004yq} and hence is constrained by observation to be small.

In the above relations $\hat S$ is the Gaussian part of the entropy field perturbation generated during inflation. The power spectrum for adiabatic and entropy field perturbations are given, at leading order, by 
\ba
\langle \zeta_i(\bfk) \zeta_i (\bfk') \rangle &=& \dfrac{2 \pi^2}{k^3} \calP_{\zeta_i} (2 \pi)^3 \delta^3 (\bfk+\bfk') 
\\
\langle \hat S(\bfk) \hat S (\bfk') \rangle &=& \dfrac{2 \pi^2}{k^3} \calP_{\hat S} (2 \pi)^3 \delta^3 (\bfk+\bfk') 
\ea
with
\ba
 \label{defA}
 \calP_{\zeta_i} &=& \calA^2 \left( \dfrac{k}{k_0} \right)^{n_{\zeta_i}-1}
 \\
\label{defB}
  \calP_{\hat S} &=& \calB^2 \left( \dfrac{k}{k_0} \right)^{n_{\hat S}-1}
\ea
where $\calA$ and $\calB$ are the amplitudes of the adiabatic and entropy modes, respectively. For simplicity, we assume approximately scale-invariant spectra for both adiabatic and entropy field perturbations, i.e. $n_{\zeta_i} \sim n_{\hat S} \sim 1$. The power spectrum for curvature and iso-curvature and the cross-power spectrum between the two are then given by \cite{Wands:2002bn} 
\ba
 \label{primordialpower}
 \calP_{\zeta} &=& \calA^2 + z_1^2 \calB^2
 \nonumber \\ 
  \calP_{\calS} &=&  s_1^2 \calB^2
 \nonumber \\
  \calC_{\zeta \calS} &=& s_1 z_1 \calB^2.
\ea
One can then define a correlation angle $\Theta$ such that
\ba 
\cos \Theta = \dfrac{\calC_{\zeta \calS}}{\sqrt{\calP_\zeta \calP_\calS}} 
\ea 
which can be written as 
\ba 
\cos \Theta = {\mathrm sgn}(z_1 s_1) \sqrt{w}
\ea 
where ${\mathrm sgn}$ is the sign function and $w$ is the fractional contribution of the entropy field perturbations, $\hat{S}$, to the curvature power spectrum
\ba
 \label{defw}
w \equiv \cos^2\Theta = \dfrac{z_1^2 \calB^2}{\calA^2 + z_1^2  \calB^2}  \,.
\ea

\begin{figure}
\includegraphics[scale=.5]{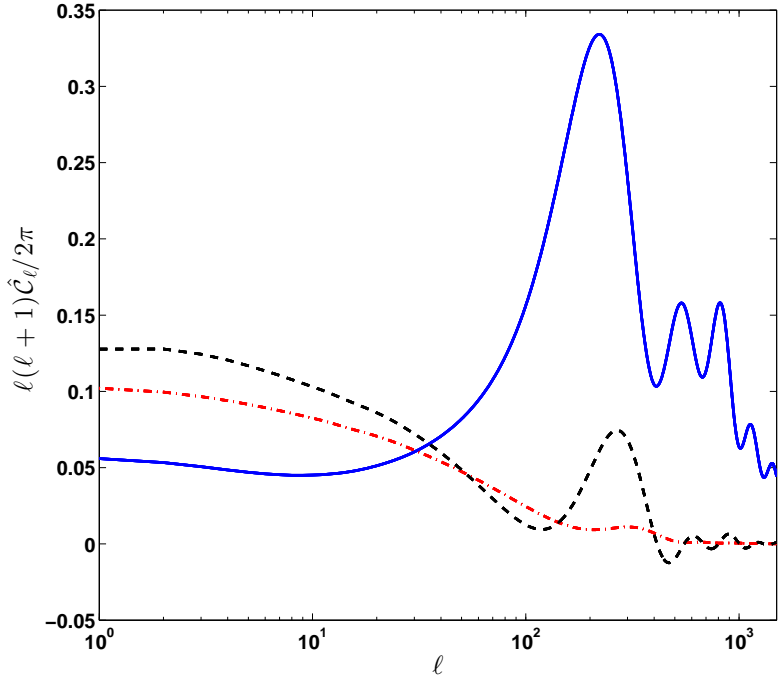}
\caption{The solid blue line is $C_\ell^{ad}$, the dashed red line is $C_\ell^{iso}$ and the dotted-dashed black line is $C_\ell^{cor}$ (the difference between power spectrum in fully correlated curvature / iso-curvature case and $C_\ell^{ad}+C_\ell^{iso}$). The cosmological parameters are $\Omega_b h^2=0.022$,$\Omega_c h^2 =0.12$ and $H_0 =67.11$.}
\label{C_ells}
\end{figure}

The CMB temperature angular power spectrum $C_\ell$ can then be written as \cite{Erickcek:2009at}
\ba
 \label{Cell}
C_\ell =  (\calA^2 +  z_1^2 \calB^2) C_\ell^{ad} +   s_1^2 \calB^2 C_\ell^{iso} +  s_1 z_1 \calB^2 C_\ell^{cor}  
\ea
where $C_\ell^{ad}$ is the adiabatic contribution to the power spectrum with $\calP_\zeta=1$, and $C_\ell^{iso}$ and $C_\ell^{cor}$ are defined analogously. Fig. \ref{C_ells} shows different contributions to $C_\ell$ obtained by CAMB code with the assumption of scale invariant power spectrum for both curvature and iso-curvature perturbations.

CMB observations are typically used to put constraints on the fractional power in the isocurvature power spectrum
\ba
\beta \equiv \dfrac{\calP_\calS}{\calP_\zeta+\calP_\calS} = \dfrac{s_1^2 \calB^2}{\calA^2 + z_1^2 \calB^2+ s_1^2 \calB^2}.
\ea
where 
\ba
w = \left(\dfrac{z_1}{s_1}\right)^2 \dfrac{\beta}{1-\beta} \,.
\ea
{}From Planck observations \cite{Ade:2013uln}, marginalising over the correlation angle we have the bound $\beta < 0.075$ at 95\% c.l.. For a curvaton model with completely correlated curvature and isocurvature perturbations ($w=1$) we have $\beta_{curv} < 0.0025$ whereas for the axion model with uncorrelated isocurvature perturbations ($w=0$) we have $\beta_{axion}<0.036$. Note, however, that for the curvaton model, the Planck group assumed that the dark matter is mainly created by curvaton decay and also that the contribution of the inflaton's perturbation to the adiabatic power spectrum, $\zeta_i$, can be neglected. 

Going beyond the leading order power spectra and considering the non-linear contributions to the curvature and isocurvature perturbations in \eqref{def_primordial} one obtains non-Gaussian primordial distributions for the curvature and isocurvature perturbations. Following \cite{Langlois:2008vk} let us define different non-Gaussianity parameters $f^{XYZ}_{NL}$ via
\ba 
\langle X_{\bfk_1} Y_{\bfk_2} Z_{\bfk_3} \rangle = (2\pi)^3 \delta({\bf k_1+k_2+k_3}) \, f^{XYZ}_{NL} 	\left[ \calP_\zeta(k_1) \calP_\zeta(k_2)+2 \, {\mathrm{ perms}}. \right]
\ea  
By the above definition and using Eqs.~(\ref{def_primordial}), (\ref{primordialpower}) and~(\ref{defw}) one can obtain
\ba 
 \label{adiabaticfNL}
f^{\zeta \zeta \zeta}_{NL} &=& \dfrac{ w^2}{z_1^2}z_2
\\
f^{\cal{S} \zeta \zeta}_{NL} &=& \dfrac{w^2}{3z_1^3} \left( s_2 z_1 + 2 s_1 z_2 \right)
\\
f^{\cal{S} \cal{S} \zeta}_{NL} &=& \dfrac{w^2 s_1}{3z_1^4} \left( 2 s_2 z_1 +  s_1 z_2 \right)
\\
f^{\cal{S} \cal{S} \cal{S}}_{NL} &=& \dfrac{s_1^2 w^2}{z_1^4}  s_2
\ea 
In the case in which $z_1=z_2=0$ the above results may not apply. Instead, one has 
\ba 
\fnl{\cal SSS} = \dfrac{\calB^4 s_1^2 s_2}{\calP_\zeta^2} = \dfrac{s_2\beta^2}{s_1^2 (1-\beta)^2} \quad \quad \quad ( z_1=z_2=0 )
\ea 
while all other three point correlations vanish.

Current CMB constraints, e.g., from Planck, only directly bound non-Gaussianity in the curvature perturbation \cite{Ade:2013ydc}
\ba
-9 < f^{\zeta \zeta \zeta}_{NL} < 15
\ea
at 95\% c.l. 
It would be interesting to see whether there is any evidence of non-Gaussian correlations in the CMB anisotropies associated with isocurvature perturbations.

\section{CMB asymmetry from non-adiabatic perturbations}

Non-linearity in the primordial curvature and isocurvature perturbations (\ref{def_primordial}) can also give rise to statistical anisotropy across our observed CMB sky in the presence of very long wavelength entropy field fluctuations during inflation.
%
To see the effect of a very long wavelength entropy mode on smaller scale CMB modes, let us decompose the Gaussian entropy field into short and long wavelength parts
\ba
\hat S = \hat S^s + \hat S^l \,.
\ea
Due to the non-linearity in Eq.~(\ref{def_primordial}), the small scale curvature and isocurvature modes are modulated by the long wavelength mode $\hat S^l$:
\ba
\label{sl-cor}
\zeta^s &\simeq & \zeta_i^s + (z_1 + z_2 \hat S^l) \hat S^s  
\\ \nonumber
{\cal S}^s_m &\simeq &  (s_1 + s_2 \hat S^l) \hat S^s 
\ea
in which we have neglected the quadratic terms in $\hat S^2$. We will see below that the correction to small scale mode due to the presence of $\hat S^l$ leads to the dipole asymmetry in CMB temperature anisotropy. On the other hand, the long wavelength perturbations are given by 
\ba
\label{long-mode}
\zeta^l &\simeq &   z_1 \hat S^l +\frac12 z_2 \hat {S^l}^2 
\\ \nonumber
{\cal S}^s_m &\simeq & s_1 \hat S^l+ \dfrac12   s_2 \hat {S^l}^2 
\ea
where we have neglected the contribution of $\zeta_i$ to the long wavelength modes. Although the wavelength of the above modes are larger than the horizon, they can have observable effects especially on low $\ell$ multipoles of CMB. We will study this effect in section \ref{lowl-constraints}.

A single long mode $\hat S^l$ can be written as a plane wave in real space
\ba
\label{long-wave}
 \hat S^l = \hat S_{k_l} \sin ({\bf k_l.x}+\varphi)
\ea
which leads to an asymmetry in the observed CMB power. In the above relation $k_l$ is the momentum of the long wavelength mode and $\varphi$ is an arbitrary phase. To maximize the effect of long mode on dipole asymmetry we set $\varphi=0$ \cite{kam1}.
Using \eqref{sl-cor} one can obtain the following correction to the observed power spectrum (\ref{Cell}) due to the long wavelength mode
\ba
\label{deltaC}
\Delta C_\ell \simeq 2 \calB^2 \hat S^l \left(   z_1 z_2  C_\ell^{ad} +    s_1 s_2  C_\ell^{iso} + ( s_2 z_1 +s_1 z_2) C_\ell^{cor} /2\right)
\ea 
It is clear from the last relation that the asymmetry originates from non-adiabaticity and non-linearity; in the absence of non-adiabatic field perturbations during inflation ($\calB=0$) or non-linearity ($z_2=s_2=0$) the asymmetry vanishes.

In what follows we mainly follow the notation as well as the procedure employed in \cite{Erickcek:2009at}. It is useful to define the fractional correction to $C_\ell$ by
\ba
\dfrac{\Delta C_\ell}{C_\ell} = 2 \hat S^l K_\ell.
\ea
where $K_\ell$  is the maximum correction one can obtain from long mode modulation (which occurs if $\hat S^l \sim 1$) and it is given by
\ba
\label{kell}
K_\ell=
\dfrac{\beta}{1-\beta}\times \dfrac{(z_1 z_2/s_1^2) C_\ell^{ad} +(s_2/s_1)C_\ell^{iso}+(z_1/2s_1)(\frac{s_2}{s_1}+\frac{z_2}{z_1})C_\ell^{cor}}{C_\ell^{ad}+ (\beta /1-\beta)\left( C_\ell^{iso} +  (z_1/s_1)C_\ell^{cor} \right)}.
\ea
Alternatively, one can rewrite the above equation by using the non-Gaussianity parameters
\ba 
K_\ell = \dfrac{z_1}{w}\times \dfrac{f^{\zeta \zeta \zeta}_{NL} C_\ell^{ad} +(z_1/s_1) f^{{\cal SSS}}_{NL} C_\ell^{iso} + \left( f^{{\cal S} \zeta \zeta}_{NL}+ (z_1/s_1)  f^{{\cal SS}  \zeta}_{NL}  \right) C_\ell^{cor}/2}{C_\ell^{ad}+(s_1/z_1)^2 \, w \, C_\ell^{iso} +  (s_1/z_1)C_\ell^{cor}}.
\ea 
The latter form is intuitively more interesting. It explicitly shows the relation between non-linearity and the dipole asymmetry. In the absence of non-Gaussianity, the dipole asymmetry vanishes as well. 

Note that in the absence of the isocurvature mode, $K_\ell$ becomes independent of scale. In this case the combination $\hat S^l K_\ell$ reduces to $A^0$ defined in \eqref{dT-anisotropy}. However in the more general case it does depend on scale and we will require this parameter to vanish (or to decrease in amplitude) for $\ell \gtrsim 64$.

Since $\beta \ll 1$ from observational constraints, we can approximate Eq. \eqref{kell} by
\ba
K_\ell \simeq
\beta\times \dfrac{(z_1 z_2/s_1^2) C_\ell^{ad} +(s_2/s_1)C_\ell^{iso}+(z_1/2s_1)(\frac{s_2}{s_1}+\frac{z_2}{z_1})C_\ell^{cor}}{C_\ell^{ad}+ \beta \left( C_\ell^{iso} +  (z_1/s_1)C_\ell^{cor} \right)} .
\ea
Furthermore, from Fig.\ref{C_ells} one can conclude that for $\beta < 0.1$ (which is the case from observations), $\beta  C_\ell^{iso}\ll  C_\ell^{ad} $ so we can further approximate by
\ba
K_\ell \simeq
\beta\times \dfrac{(z_1 z_2/s_1^2) C_\ell^{ad} +(s_2/s_1)C_\ell^{iso}+(z_1/2s_1)(\frac{s_2}{s_1}+\frac{z_2}{z_1})C_\ell^{cor}}{C_\ell^{ad}+  \beta  (z_1/s_1)C_\ell^{cor} } .
\ea

Since the observational bounds on the hemispherical asymmetry has been put under the assumption of a scale invariant asymmetry, we need to average over $K_\ell$'s to obtain such effective amplitude over all independent modes from $\ell_{min}=2$ to $\ell_{max}=64$ at which range the asymmetry has been observed. That is, we can define
\ba
\label{A-effective}
A \equiv \hat S^l \sum_{\ell_{min}}^{\ell_{max}} \dfrac{(2\ell+1)}{(1+\ell_{max}-\ell_{min})(1+\ell_{max}+\ell_{min})} K_\ell 
\equiv \hat S^l \tilde{A}. 
\ea
In the case of scale independent asymmetry one can check that $A=\hat S^l K_\ell = A^0$ where $A^0$ is defined in \eqref{dT-anisotropy}. Since $\hat S^l <1$, $\tilde{A}$ is the largest amplitude of asymmetry which can be generated in the model. To satisfy the observed amplitude of asymmetry we require $A \simeq 0.07$ which means we need to have  
$\tilde{A} \gtrsim 0.07$. At the same time, we need $K_\ell$ to decay to much smaller values for $\ell > 64$ to satisfy the absence of dipole asymmetry observation at smaller scales. We will check these constraints for two specific models, namely, the curvaton and axion models of dark matter isocurvature. To see explicitly that the dipole asymmetry will decay away at smaller scales we also compute the effective amplitude of asymmetry at smaller scales, i.e. for $600 < \ell < 1500$ and show that it would be much smaller than the asymmetry at largest scales obtained in Eq. \eqref{A-effective}. Note that the above formalism is quite independent of the model and can be applied to many other models of isocurvature. 


\section{Low multipole constraints on long-wavelength modes}
\label{lowl-constraints}
As we mentioned in the previous section the long wavelength mode can also affect the low $\ell$ multipoles on CMB anisotropy besides its effect on dipole asymmetry. Requiring that this effect is small enough to be consistent with observations puts several constraints on model parameters. In this section we investigate such constraints. We follow the notation in \cite{kam2} but will generalize it in a model independent manner. Let us start by expanding the large scale gravitational potential in orders of ${\bf(k.x)}$
\ba 
\Psi^l(k_l,x) = \Psi_{k_l} \sum_{n=0}^\infty \alpha_n ({\bf k_l.x})^n
\ea   
This long wavelength gravitational potential is related to the primordial long wavelength perturbations obtained in Eq. \eqref{long-mode}. As a results this parameter is a combination of adiabatic and isocurvature modes. We will make this relation explicit below. Note that different powers of $({\bf k_l.x})$ appear by replacing long wavelength modes in \eqref{long-mode} by the explicit expression Eq. \eqref{long-wave} and expanding the results in powers of $({\bf k_l.x})$. 

On the other hand,
the above gravitational potential is related to the temperature anisotropy by the transfer functions
\ba 
\dfrac{\Delta T}{T} (k_l,x)= \Psi_{k_l} \sum_{n=0}^\infty \alpha_n T_n ({\bf k_l.x})^n. 
\ea  
The transfer coefficients, $T_n$, up to $n=3$ have been computed in \cite{kam2}. The multipole moments in temperature anisotropy are then given by
\ba 
a_{\ell m} = \int \left( \dfrac{\Delta T}{T} \right) \, Y^*_{\ell m} (\theta,\phi) d\Omega = 
\Psi_{k_l} \int \left( \sum \alpha_n T_n (k_l x)^n \cos^n \theta \right) \, Y^*_{\ell m} (\theta,\phi) d\Omega
\ea 
in which we have used ${\bf k_l.x}=k_l x \cos \theta$. Since in this expansion we have no $\phi$ dependence, this particular form of anisotropy only contributes to $m=0$ multipoles. Hence we have 
\ba 
a_{\ell 0} =
\Psi_{k_l} \int \left( \sum \alpha_n T_n (k_l x)^n \cos^n \theta \right) \, Y^*_{\ell 0} (\theta,\phi) d\Omega.
\ea 
In the above integral, for each $\ell$ the leading order non-zero contribution comes from the term $n=\ell$ for which we can effectively replace $\cos^n \theta$ by its corresponding spherical harmonic
\ba 
\cos^n \theta \to 2^n \dfrac{(n !)^2}{2n!} \sqrt{\dfrac{4 \pi}{2n+1}} Y_{n0}.
\ea 
Plugging this back into the integral and using the ortho-normality conditions of spherical harmonics one can obtain
\ba 
a_{\ell 0}  \simeq \Psi_{k_l} \, \alpha_\ell \, T_\ell \, ({k_l} x)^\ell \, \dfrac{2^\ell (\ell!)^2}{2\ell!}  \sqrt{\dfrac{4 \pi}{2\ell+1}}.
\ea
Now we can use this formalism to compute multipoles in our model. Note that the transfer function is different for curvature and iso-curvature perturbations, so we may decompose $\Psi$ into two parts
\ba 
\Psi^{l,ad} &=& -\dfrac{3}{5} \zeta^l = -\dfrac{3}{5} \left( z_1 \hat S_{k_l} ({\bf k_l.x}) + \dfrac{1}{2} z_2 \hat S_{k_l}^2 ({\bf k_l.x})^2 - \frac16 z_1 \hat S_{k_l} ({\bf k_l.x})^3 +...\right) 
\\
\Psi^{l,iso} &=& -\dfrac{1}{5} \dfrac{\Omega_{cdm}}{\Omega_{cdm}+\Omega_b} {\cal S}^l_m = -\dfrac{1}{5} \dfrac{\Omega_{cdm}}{\Omega_{cdm}+\Omega_b}  \left( s_1 \hat S_{k_l} ({\bf k_l.x}) + \dfrac{1}{2} s_2 \hat S_{k_l}^2 ({\bf k_l.x})^2 - \frac16 s_1 \hat S_{k_l} ({\bf k_l.x})^3 +...\right) 
\ea 
As for the transfer functions, we note that the only difference is in the SW effect while ISW and Doppler effects  would be similar for the adiabatic and iso-curvature initial conditions. Hence we have 
\ba 
T_\ell^{iso} = T_\ell^{ad} + \dfrac{5}{3}
\ea 
Furthermore, it has been shown in \cite{kam2} that the SW, ISW and Doppler effects cancel each other for $\ell =1$ for adiabatic initial condition, so we have $T_1^{ad} =0$ and $T_1^{iso}= 5/3$. We also have $T_2^{ad} \simeq T_3^{ad} \simeq 0.3$. Then, one can obtain the following model independent expressions for low $\ell $ multipoles: 
\ba 
a_{10} &\simeq& -\dfrac23 \sqrt{\dfrac{\pi}{3}} \dfrac{\Omega_{cdm}}{\Omega_{cdm}+\Omega_b}  s_1 \hat S_{k_l} (k_l x) 
\\
a_{20} &\simeq& - \dfrac23 \sqrt{\dfrac{\pi}{5}} \left[3 z_2 T_2^{ad}+  \dfrac{\Omega_{cdm}}{\Omega_{cdm}+\Omega_b}  s_2 T_2^{iso} \right] \hat S_{k_l}^2 (k_l x)^2 
\\
a_{30} &\simeq&  \dfrac{2}{75} \sqrt{\dfrac{\pi}{7}} \left[3 z_1 T_3^{ad}+  \dfrac{\Omega_{cdm}}{\Omega_{cdm}+\Omega_b}  s_1 T_3^{iso} \right] \hat S_{k_l} (k_l x)^3 
\ea 
To compare the above results to the observations we roughly estimate the upper bound by $a_{\ell 0} \lesssim 3 \sqrt{C_\ell}$, three times the rms value of each multipole. Recent Planck observations then put the following upper bounds \cite{Ade:2013zuv,Planck-website} 

\ba 
a_{10} \lesssim  3.6 \times 10^{-3} \quad , \quad 
a_{20} \lesssim 1.9 \times 10^{-5}  \quad , \quad 
a_{30} \lesssim 2.5 \times 10^{-5}.
\ea 
From constraints on dipole and quadrupole we have 
\ba 
s_1 \hat S_{k_l} (k_l x) \lesssim 5.8 \times 10^{-3}
\\
\left[ z_2 + 1.7 s_2 \right]  \hat S_{k_l}^2 (k_l x)^2 \lesssim 3.6 \times 10^{-5} 
\ea 
In the presence of non-linearity, the constraints from octupole are subleading.

\section{Asymmetry in curvaton model}

We can now apply the above model independent results to specific example of a mixed inflaton+curvaton model. The source of the non-adiabatic perturbation during inflation is the curvaton field
\ba
\hat S &=& 2 \dfrac{\delta \sigma}{\sigma} \,,
\ea
in addition to the adiabatic, inflaton field perturbations, $\zeta_i=-H\delta\phi/\dot\phi$. 
We then have in Eqs.(\ref{defA}) and~(\ref{defB})
\ba
\calA^2 &=& \dfrac{H_{inf}^2}{8\pi^2 \mPl^2 \epsilon_{inf}}
\\
\calB^2 &=& \dfrac{H_{inf}^2}{\pi^2 \sigma^2}.
\ea
The coefficients of the linear and non-linear contributions to the primordial density from curvaton fluctuations in (\ref{def_primordial}) are given in the sudden-decay approximation by \cite{Langlois:2008vk,Fonseca:2012cj,Sasaki:2006kq} 
\ba
z_1 &=& \dfrac{R}{3} 
\\
\label{z2express}
z_2 &=& \dfrac{R}{9} \, \left(\dfrac{3}{2}- 2 R -R^2 \right) 
\ea
in which $R$ is the curvaton fractional energy density defined by 
\ba
R = \dfrac{3 \rho_\sigma}{3 \rho_\sigma + 4 \rho_\gamma}
\ea
which yields 
\ba 
\fnl{\zeta \zeta \zeta} = \dfrac{w^2}{R} \left( \dfrac{3}{2} -2R - R^2  \right)
\ea 
Note that our expression (\ref{z2express}) for the second-order coefficient, $z_2$, includes the full dependence of $R$ and hence $N$ on the curvaton field value, $\sigma$. Our expression reduces to that given in \cite{Erickcek:2009at} for $R\ll1$ but is more general and remains valid for $R\sim1$ \cite{Fonseca:2012cj}. There are several terms which start from second order in $R$ where this correction is important.

The general relations for the coefficients for the linear and non-linear matter isocurvature perturbations, $s_1$ and $s_2$ in (\ref{def_primordial}), are model dependent and we present here only two limiting cases.

\subsubsection*{Case I: CDM is created from curvaton decay}
In this case we have \cite{Langlois:2008vk} 
\ba
\label{case1s1}
 s_1 &=& 1-R
 \\
  s_2 &=& - \dfrac{(1-R)}{6} (3+6 R +2 R^2)
\ea 
which yields
\ba 
\fnl{\cal S S S} = -\dfrac{27 w^2}{2 R^4} (1-R)^3 (3+6 R+2 R^2)
\\
\fnl{{\cal S  S}\zeta} = - \dfrac{3 w^2}{2R^3} (1-R)^2 (3+16R+6R^2)
\\
\fnl{{\cal S}\zeta \zeta} = - \dfrac{w^2}{2R^2} (1-R) (-3+14R+6R^2)
\ea 

The dipole and quadrupole constraints simplify to 
\ba 
(1-R) \dfrac{\delta \sigma}{\sigma}(k_lx) \lesssim 2.9 \times 10^{-3}
\\
(0.85 + 0.7 R -0.9 R^2 -0.5 R^3) \left(\dfrac{\delta \sigma}{\sigma}\right)^2 (k_l x)^2 \lesssim 9.5 \times 10^{-6}
\ea 
Interestingly, the dipole constraint is automatically satisfied in the most interesting case, i.e. in the limit $R \to 1$ while to satisfy quadrupole constraint we require
\ba 
 \label{case1bound}
\left(\dfrac{\delta \sigma}{\sigma}\right) (k_lx) \lesssim 8 \times 10^{-3}.
\ea 

\begin{figure}
\includegraphics[scale=.5]{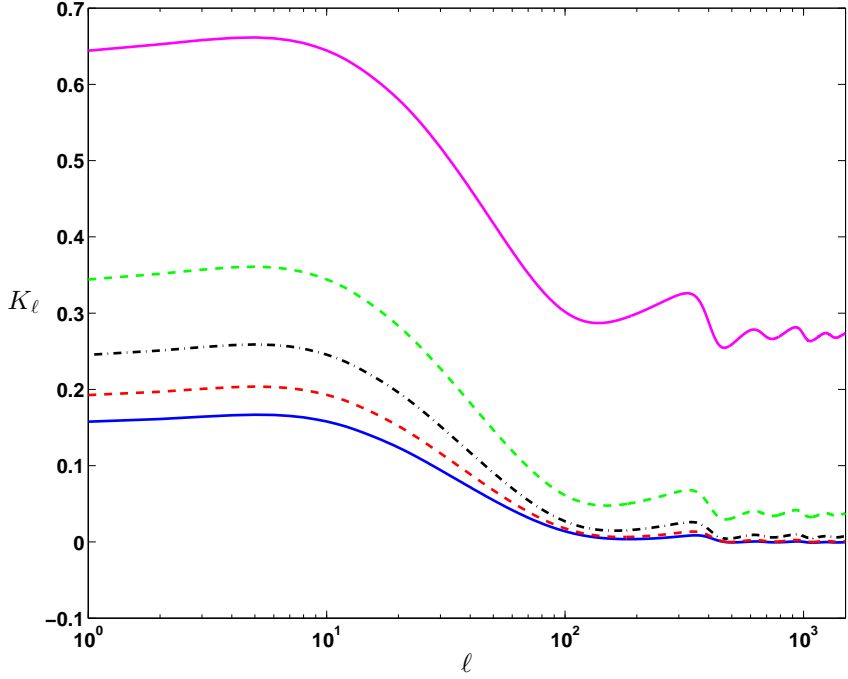}
\caption{$K_\ell$ versus $\ell$ when dark matter is created from curvaton decay with $\beta =0.075$. The plots has been made for different values of $R$. From top to bottom we have set $R=0.9, 0.8, 0.7, 0.6, 0.5$ with the effective amplitude of asymmetry $\tilde{A}= 0.38, 0.12, 0.07, 0.05, 0.04$, respectively.}
\label{Kl1}
\end{figure}

\begin{figure}
\includegraphics[scale=.5]{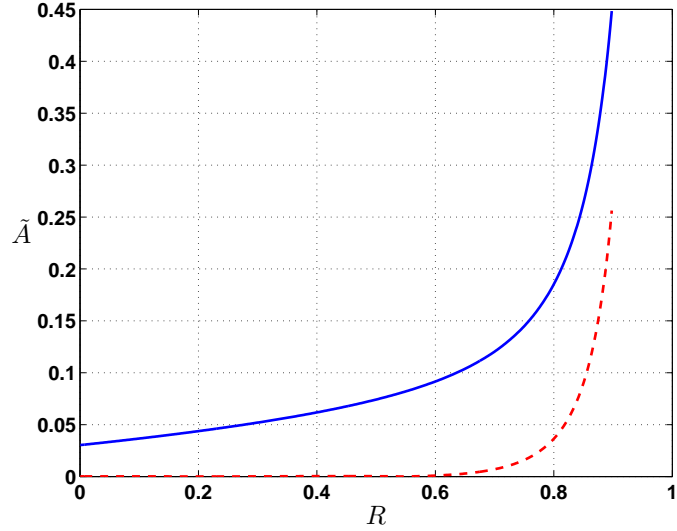}
\caption{The maximal asymmetry parameter $\tilde{A}$ defined in Eq.~\eqref{A-effective} at large scales (solid blue line, $2<\ell <64$) and smaller scales (dashed red line, $600 < \ell < 1500$) as a function of $R$ for the case in which dark matter is created from curvaton decay. We have set $\beta =0.075$.}
\label{tildeA1}
\end{figure}

The observed hemispherical asymmetry is given in Eq.~\eqref{A-effective} as $A=\tilde{A}\hat S^l$ where for the curvaton we have $\hat S^l \sim 2 (\delta \sigma/\sigma) ({\bf k_l .x})$. Thus to obtain the observed CMB asymmetry for low multipoles, $A\sim0.07$, given the upper bound (\ref{case1bound}) we require $\tilde{A} \gtrsim 4$. As shown in figure \ref{tildeA1}, this large value for $\tilde{A}$ can be obtained only as $R$ approaches unity (its maximum value), and in fact we require $R>0.97$ to obtain the observed CMB asymmetry at low multipoles. However, in the limit $R \to 1$ we cannot get any strong scale dependence in the amplitude of asymmetry, as is clear from figure~\ref{Kl1}. This is because in the limit $R\to1$ the matter isocurvature perturbation vanishes [$s_1\to0$ in Eq.~(\ref{case1s1})] and the whole asymmetry comes from the primordial adiabatic density perturbation, for which we already know the asymmetry does not decay at smaller scales.


\subsubsection*{Case II: Curvaton decay has negligible effect on dark matter}

\begin{figure}
\includegraphics[scale=.5]{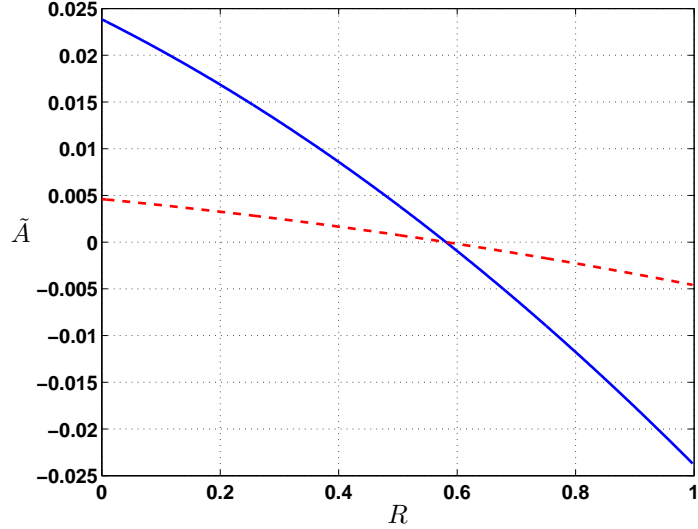}
\caption{The maximal asymmetry parameter $\tilde{A}$ defined in Eq.~\eqref{A-effective} at large scales (solid blue line, $2<\ell<64$) and smaller scales (dashed red line, $600 <\ell < 1500$) as a function of $R$ for the case in which the effect of curvaton decay on dark matter is negligible with $\beta =0.075$.}
\label{tildeA2}
\end{figure}

If the dark matter abundance is created well before the curvaton decays then we have  \cite{Langlois:2008vk} 
\ba
 s_1 &=& -R
 \\
  s_2 &=& - \dfrac{R}{3} \, \left(\dfrac{3}{2}-2 R - R^2 \right)
\ea
which yields 
\ba 
\fnl{\cal S S S}=-3\fnl{{\cal S S}\zeta}=9 \fnl{{\cal S}\zeta \zeta}=-27 \fnl{\zeta \zeta \zeta}
\ea 
and $K_\ell$ simplifies to
\ba
K_\ell \simeq
\dfrac{\beta}{27}\, \left(\dfrac{3}{2}-2R-R^2\right)\times \left( \dfrac{ C_\ell^{ad} +9 C_\ell^{iso}-3C_\ell^{cor}}{C_\ell^{ad}+  \beta  C_\ell^{cor}/3 }\right) .
\ea
Note that the last two factors are not typically larger than one so one can estimate
\ba
K_\ell \lesssim 10^{-1} \beta < 10^{-2}.
\ea
$K_\ell$ in this case, as a function of $R$ is maximum in the limit $R\to 0$ or $R \to 1$.
The dipole and quadrupole constraints give 
\ba 
R \left(\dfrac{\delta \sigma}{\sigma}\right) (k_l x) \lesssim 2.85 \times 10^{-3}
\\
R  \, \left\vert -0.7 + 0.9 R+0.46 R^2 \right\vert \, \left(\dfrac{\delta \sigma}{\sigma}\right)^2 (k_lx)^2 \lesssim 9.5 \times 10^{-6}
\ea 
By reducing $R$ one can easily satisfy the above constraints, though one has to be also careful about the constraints on the amplitude of local non-Gaussianity. However, as it is clear from figure \ref{tildeA2}, it seems the amplitude of asymmetry is too small in this limiting cases to be able to explain the observed asymmetry. 

\section{Axion model}

In axion model we have $z_1 = z_2 =0$ since axion does not decay into relativistic particles and hence does not provide a source for the primordial adiabatic density perturbation. This is an interesting property of the axion model which has an important consequence for the asymmetry; the isocurvature mode is responsible for the whole asymmetry so the asymmetry is automatically scale dependent in the CMB and can be made consistent with observations.
 
The dark matter isocurvature perturbation in this model is given by \cite{Kawasaki:2008sn}
\ba 
\calS = r \left[ \dfrac{2 a_i \delta a}{a_*^2} + \left( \dfrac{\delta a }{a_*} \right)^2 \right]
\ea 
in which $r = \Omega_a/\Omega_{DM}$ is the ratio of the axion abundance to the total dark matter, $\delta \sigma$ is the perturbation of the axion field and we have defined 
\ba 
a_i = F_a \theta 
\\
a_* = {\mathrm max} \left\{ F_a \theta \, , \dfrac{H_{inf}}{2 \pi}  \right\}
\ea 
Here $F_a$ is the energy scale of  Peccei-Quinn symmetry breaking and $\theta$ is the initial misalignment angle of the axion.  Note that the above isocurvature is compatible with observations in the limit $r \ll 1$. 

Since we are not interested in strongly non-Gaussian models we assume $F_a \theta > H_{inf}/2\pi$. In this case we have 
\ba 
s_1 = s_2 =2 r 
\ea 
and 
\ba 
\calC_\ell  &=& \calA^2  \calC_\ell^{ad} + s_1^2 \calB^2  \calC_\ell^{iso} 
\\
\calB  &= & \dfrac{H_{inf}}{2 \pi a_i}
\\
\label{axionfNLSSS}
\fnl{\cal S S S} &=& \dfrac{ \beta^2}{2r(1-\beta)^2}
\ea 
and all other non-Gaussianity parameters vanish.

The scale dependent asymmetry $K_\ell$ is then given by
\ba 
\label{axionKl}
K_\ell = \dfrac{\beta}{1-\beta} \times \dfrac{ \calC_\ell^{iso} }{\calC_\ell^{ad} + \beta \calC_\ell^{iso}/(1-\beta)}
\ea 
Interestingly, $K_\ell$ only depends on $\beta$.  In the limit at which $\beta \ll 1$ as well as $\beta \calC_\ell^{iso} \ll \calC_\ell^{ad}$ this simplifies to 
\ba 
\label{axionKl2}
K_\ell \simeq \beta \times \dfrac{\calC_\ell^{iso}}{\calC_\ell^{ad}}.
\ea 
\begin{figure}
\includegraphics[scale=.5]{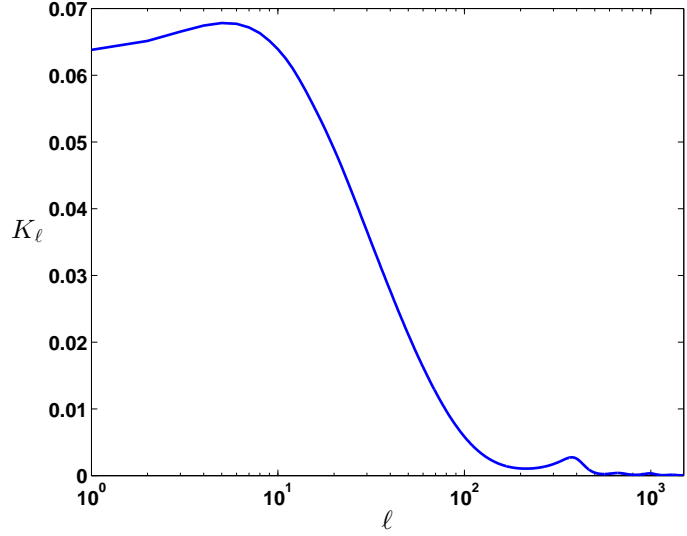}
\caption{A plot for $K_\ell$ for axion model. In this plot we have set $\beta = 0.036$ which yields to $\tilde{A} = 0.0283$. }
\end{figure}
As for the constraints from low $\ell$ multipoles, one can easily check that the strongest constraint comes from quadrupole which simplifies to 
\ba 
\label{axionquad}
 r^{1/2} \, \left(\dfrac{\delta a }{a}\right) (k_l x) \lesssim 3 \times 10^{-3}.
\ea 
The interesting point is that $r$ appears here in this constraint while it does not appear in the expression (\ref{axionKl}) for hemispherical asymmetry. Hence, one can reduce $r$ in order to satisfy the quadrupole constraint while the asymmetry (\ref{axionKl}) remains unchanged. 

For axion isocurvature perturbations (with $\beta\ll1$) uncorrelated with the adiabatic density perturbation we find, from Eq.~\eqref{A-effective}, the maximal asymmetry on large scales
\ba
\label{axionA}
\tilde{A} \simeq 0.806 \beta \quad {\rm for}\ 2< \ell< 64 \,,
\ea
while on smaller angular scales we find
\ba
\tilde{A} \simeq 0.0042 \beta \quad {\rm for}\ 600 < \ell< 1500 \,.
\ea

Imposing the tight observational bound on the isocurvature fraction, $\beta<0.036$, requires the asymmetry even on large scales (\ref{axionA}) to be small, $\tilde{A}<0.03$. However one should bear in mind that this bound is obtained assuming isotropy (i.e., no asymmetry). If the constraints on $\beta$ in an anisotropic model are relaxed then it may be possible to obtain a larger asymmetry on large scales, closer to the observed value $A\sim0.07$, while leaving the CMB isotropic on smaller angular scales.

Finally, we note that although the adiabatic non-Gaussianity parameter, $f_{NL}^{\zeta\zeta\zeta}$ in Eq.~(\ref{adiabaticfNL}) automatically vanishes in the axion case since $z_2=0$,  in order to obtain a maximal value for the axion density perturbation, $\hat{S}^l\sim1$, we require $r\lesssim10^{-5}$ in Eq.~(\ref{axionquad}) and hence a strongly non-Gaussian matter isocurvature perturbation, $\fnl{\cal S S S}\gtrsim 10^5\beta^2$ in Eq.~(\ref{axionfNLSSS}).

\section{Conclusions}

In this work we 
 presented a general model-independent formalism relating CMB dipolar asymmetry to matter isocurvature perturbations. We have shown that non-linearity due to non-adiabatic perturbations during inflation can lead to a dipole asymmetry due to the correlation between a very long wavelength (super-horizon) modes today and CMB scales.  The relation between non-linearity and asymmetry can be easily understood. Any observable effect of a very long mode on shorter wavelengths cannot be generated at first-order level simply because different wavelengths are decoupled in the linear regime. However, at non-linear order the very long mode can interact with shorter modes leaving observable imprints on them. Hence there should be a relation between the strength of the non-linear interaction (i.e., the amplitude of non-Gaussianity) and the dipole asymmetry. 
See, for example, Ref.~\cite{Namjoo:2014nra} for more details and discussion about the relation between non-Gaussianity (in the squeezed limit) and dipole asymmetry.    
 
We showed this relation explicitly in this work through our formalism, based on the generalised $\delta N$ formalism \cite{Langlois:2008vk}, which is independent of the specific model and can be applied to any model of dark matter (or baryon) isocurvature fluctuations as well as the primordial adiabatic density perturbation.
Besides obtaining general relations between non-linearity and the dipole asymmetry, we derived formulae for the effect of large scale entropy perturbations on the low $\ell$ multipoles of the CMB. We also updated constraints from this effect on models of dipole asymmetry due to long mode modulations using recent Planck data. 

We studied two well-known examples for generating matter isocurvature perturbations: the curvaton and axion models. We showed that for the curvaton model, producing isocurvature perturbations correlated with the adiabatic density perturbations at first order, it is difficult to satisfy all observational constraints while, at the same time, generating a significant dipole asymmetry on large angular scales. For the axion model, however, we showed that satisfying observational constraints is much easier when the isocurvature mode is un-correlated to the adiabatic mode. On the other hand, the difficulty in this case comes from constraints on the allowed amplitude of matter isocurvature perturbations which implies that they can only marginally generate the observed asymmetry. 

It is worth to compare our results with the ones obtained in \cite{Erickcek:2009at}. In that paper, only the curvaton model has been considered. While our results holds for the whole range of $R$ in curvaton model,  their results are reliable only in the non-Gaussian adiabatic limit, i.e. for $R\ll 1$. Hence, the two results match at leading (linear) order in $R$. They, however, considered the case in which the curvaton decays has a small but non-zero effect on dark matter. This is somewhat similar to the axion model in the limit $r \ll 1$ for which we can get rather large dipole asymmetry. 

We note that existing constraints on the amplitude of isocurvature perturbations assume an isotropic distribution across the CMB sky. These constraints may no longer be reliable in the presence of a dipole asymmetry. It would be interesting to consider a generalised analysis allowing for a dipolar asymmetry, especially after the next Planck data release including polarisation data.

We have considered the effect of a non-zero bispectrum in the squeezed limit, correlating short wavelength fluctuations with very long wavelength modes leading to an asymmetry in the power spectrum observed on the CMB sky. We note that at higher order, the effect of a non-zero trispectrum in the squeezed limit can lead to an inhomogeneous bispectrum \cite{Byrnes:2011ri} and hence the possibility of asymmetry in the bispectrum being observed on the CMB sky. We await to see if the final release of Planck data can shed new light on the wide range of asymmetries possible in the observed CMB in the presence of very long, super-horizon perturbations .

\section*{Acknowledgment }
We thank Ali Akbar Abolhasani, Yashar Akrami, Shant Baghram, Razieh Emami and Moslem Zarei for fruitful discussions and correspondences. The work of MHN is supported in part by a NSF grant PHY-1417421.
This work was supported by STFC grants ST/K00090X/1 and ST/L005573/1.
	
{}

\end{document}